# Angular momentum balance and vortex production in wall-bounded flows *

**A. Paglietti**

*University of Cagliari, 09100 Cagliari, Italy*
*E-mail: paglietti@unica.it*

**ABSTRACT.** To produce a vortex, a torque must be applied to the fluid. In viscous fluids, the torques that produce turbulent vortices result from the loss of symmetry of the stress tensor, once the viscous friction exceeds the shear stress resistance of the fluid. In wall-bounded flows, in particular, the turbulent vortices form in a thin layer of fluid adjacent to the wall, practically coinciding with the so-called viscous sublayer, where the viscous friction reaches the largest values. The present paper determines a vortex structure for this sublayer, consistent with the well-known linearity of the diagram of the mean streamwise velocity of this region. The analysis enables us to calculate the diameter, angular velocity, and interaxis of the vortices in the viscous sublayer in steady-state conditions. The lifting force that makes the vortices migrate from the wall towards the mainstream flow is determined, and the crucial role played by gyroscopic precession in the reorientation of the vortex axis is discussed.

# I. INTRODUCTION

Turbulence of viscous fluids in wall-bounded flows is the primary concern of the present paper. The phenomenon is quite common in practical applications. It ranges from fluid flow over a plate (plane or curved) to channel and pipe flow. The paper inquires about the relationship between the origin of turbulence and the friction shear stress acting on the fluid during the flow. In wall-bounded flows, turbulence is produced near the walls, where the shear stress is largest. Turbulence can also occur in the absence of bounding walls, as usually happens to flows containing fluid jets and plumes. The details of these cases are not considered here. The present analysis, however, should suggest how similar results can be extended to more general situations.

Generally speaking, the flow of a viscous fluid becomes turbulent due to the formation of a multitude of small vortices that move randomly within the fluid, still keeping pace with the average velocity of the flow. These vortices will henceforth be referred to as *turbulent vortices*. They consist of one or more axially-symmetric segments of fluid rotating about their symmetry axis. Each vortex segment has its spin angular momentum (or *spin*, for brevity) relative to the segment's centre of mass or — which is equivalent under axial symmetry conditions — relative to the segment's axis. Spin is an intrinsic property of a moving body because its value is independent of the motion of the centre of mass. This property distinguishes the spin from the *orbital* angular momentum. The latter is the angular momentum due to the velocity of the centre of mass; it depends both on the velocity of the inertial frame of reference in which the motion is described and the point relative to which it is calculated.

Turbulent vortices, and thus their spin, are produced when the flow velocity exceeds a given limit. The higher the velocity is above this limit, the greater the amount of vortical spin produced. Thus, the specific spin content of a fluid depends on the average velocity of the flow or, more precisely, on the flow's Reynolds number. However, a torque needs to be applied to the fluid that makes up a vortex to make it spin. Though seldom acknowledged in fluid dynamics, this is the inescapable consequence of the angular momentum balance law. As discussed in Sect. **II**, no turbulent vortex can exist in a portion of fluid if no torque has ever been applied to that portion.

The traditional (non-polar) approach to viscous fluids does not contemplate the presence of spin-producing torques in the fluid. In that approach, the angular momentum balance equation reduces to the mere requirement that the antisymmetric part of the stress tensor should vanish, which is a condition that is usually trivially met since the stress tensor of a non-polar material is assumed to be symmetric by constitutive hypothesis. Under these conditions, the angular momentum balance has no active role in controlling the motion of the fluid particles, which makes classical fluid mechanics inherently incapable of dealing with the torques that must arise locally in a fluid to produce a turbulent vortex.

So, how can a torque ever be produced in a given volume of a classical viscous fluid and a turbulent vortex generated? Section **III** answers this question. The viscous friction forces act on surfaces that are parallel to the streamlines. These forces can assume any value, no matter how large, provided that the velocity gradient in the normal direction to the streamlines is kept sufficiently large. There are no viscous forces, however, on the surfaces normal to the streamlines, simply because there is no flow over these surfaces. On these surfaces, the shear stress component cannot exceed the shear stress resistance of the material. If the applied viscous friction exceeds the shear stress limit of the material, the stress tensor ceases to be symmetric. When this happens, the resultant torque of the shear stress acting on the element surface is no longer zero and it drives the element rotation.

Essential to the arguments that follow is a clear distinction between rotation and vorticity. The vorticity $\omega$ at a point of a fluid is the curl of the velocity field at that point. That is, $\omega = \nabla \times \mathbf{v}$, where $\mathbf{v}$ denotes the velocity vector at the points of the fluid. The vorticity is known to be equal to twice the angular velocity vector of a small solid sphere immersed in the fluid and flowing with it (cf., e.g., [1, 2]). The sphere's rotation, however, is produced by the different velocity of the fluid particles in contact with its surface. It does in no way imply a rotation of the fluid that corresponds to the sphere location. In fact, the vorticity may be different from zero, which would make the sphere rotate, and yet the fluid possesses no rotation at all. This is what happens, for instance, in the laminar flow through a rectilinear pipe and, more generally, in any non-uniform laminar flow with rectilinear pathlines. Conversely, the vorticity may be zero throughout the fluid, and yet the fluid rotates about a given axis, as in the classical case of the *irrotational* vortices (cf., e.g., [3]). In short, vorticity and rotation are two independent properties of the particles of a flowing fluid.

The distinction between vorticity and rotation has important implications on the turbulent vortices and hence the vortex segments. As observed in Robinson [4], a vortex need not follow the Helmholtz laws for the vorticity lines. Thus, the portion of rotating fluid making up a vortex may abruptly end up within the fluid and may move at a velocity that is different, both in value and in direction, from the velocity of the surrounding fluid. This is at odds with what applies to the vorticity lines in the inviscid limit.

Of course, to approach fluid dynamics in terms of vorticity [2] does not solve the problem of finding the torques needed to produce turbulent vortices. The vorticity approach is based on the so-called vorticity equation, which is nothing but the curl of the linear momentum balance equation [1] and which, therefore, is equivalent to it [5], [6]. Because no torque is involved in the linear momentum balance equation, also the vorticity equation-based approach is incapable, as the traditional approach, to deal with spin-producing torques and, thus, turbulent vortex production.

Resorting to flow instability does not solve the problem of determining the origin of turbulent vortices. The instability of a solution of the classical fluid dynamic equations means that the solution changes dramatically with the slightest change in the initial-boundary conditions. This may well explain why the flow becomes unsteady and even chaotic when the flow's Reynolds number exceeds the value beyond which the flow becomes unstable. The chaotic character of the flow, however, is quite another thing from its content of turbulent vortices. The classical fluid dynamic equations ignore the possibility that, under appropriate conditions, vortex-producing torques may be acting on the fluid. This at the outset excludes turbulent vortex formation from the theory, no matter the instability of the flow. It is not surprising, then, that flow instability fails to explain why turbulence arises even in unconditionally stable flows.

The numerical simulation of turbulence in fluid flows is not exempt from the same shortcoming discussed above, as it is based on the equations of classical fluid dynamics. In order to model turbulence, all the simulations seed – in one way or another– some germ of turbulence into the computational procedure, quite often at the boundary of the computing domain. Wu [7] provides an updated review of the various methods adopted for this purpose. Under these conditions, the numerical procedure cannot produce a genuine, new insight into the origin of turbulence. Moreover, the lack of a clear-cut distinction between rotation and vorticity often hampers the correct interpretation of the obtained results [4].

By applying the analysis of Sect. **III** and still remaining in the realm of a classical viscous fluid, the problem of the vortex producing torques in wall-bounded flows is considered in Sect. **IV**. We show that the vortex producing torques form in a thin zone near the wall, practically coinciding with the so-called viscous sublayer. This zone only exists if the shear stress at the wall exceeds the shear stress limit of the material.

The presence of the viscous sublayer in a turbulent near-wall flow is a well-established experimental fact. We know that, in general, this sublayer becomes thinner and thinner as the flow velocity increases and that, moreover, the average streamwise velocity of its particles is a linear function of the distance of the particles from the wall. However, the flow in the same sublayer is far from laminar (cf., e.g., [4] and the other references quoted in Sect. **IV**).

Section **V** shows that the linearity of the average streamwise velocity of the viscous sublayer is compatible with the presence of vortices in it. The vortices are created by the roll-up of the fluid layer near the wall [8] as its shear stress limit is exceeded. Circular cylindrical transverse vortices are thus formed, which roll rigidly on the wall, without slipping. Their motion produces a streamwise component of the fluid particles velocity that varies linearly with the distance from the wall and is equal to zero on the wall, which is in agreement with the no-slip condition. The analysis provides the formulae to calculate the diameter and the angular velocity of the vortices at their formation. It also determines the admissible range of the vortex interaxis in the viscous sublayer at steady-state.

Finally, Sect. **VI** discusses how the rolling vortices can leave the wall to migrate outside of the viscous sublayer, thus going on to populate the whole flow with turbulent vortices. The Magnus effect drives vortex lifting. The lift force per unit vortex length is calculated and compared to the vortex weight. This section also shows how small irregularities in the flow may produce a substantial reorientation of the vortex axis, both in a streamwise plane and in the transverse plane. This phenomenon appears to be responsible for the wide variety of vortex orientations which is characteristic of all turbulent flows. It is essentially a gyroscopic precession effect, made possible by the comparatively large spin angular momentum of these vortices.

Besides offering a new perspective on the origin of turbulence, the analysis presented in this paper should provide the basis to improve the modelling of turbulent flow in viscous fluids. Of course, a turbulent flow may manifest a number of other irregularities, such as puffs and slugs, which are different than turbulent vortices and free from spin angular momentum. The present analysis does not concern them. Moreover, no attempt is made to extend to polar fluids the limit stress approach to turbulence considered here.

## II. RESULTS FROM CLASSICAL MECHANICS

For systems of particles or continuous bodies, the angular momentum $\boldsymbol{L}$, relative to a given point $O$, can always be decomposed as follows (cf., e.g., [9-11] and for continuous bodies, [12]):

$$\boldsymbol{L} = \boldsymbol{L}_{\text{orb}} + \boldsymbol{L}' . \tag{1}$$

The quantity $\boldsymbol{L}_{\text{orb}}$ is the orbital angular momentum of the system. It is defined as the angular momentum of the total mass $M$ of the system, assumed to be concentrated at the system's centre of mass $C$ and moving with it. That is

$$\boldsymbol{L}_{\text{orb}} = \mathbf{c} \times M \dot{\mathbf{c}} , \tag{2}$$

where $\mathbf{c}$ is the position vector of the centre of mass relative to point $O$, see Fig. 1. The other quantity $\boldsymbol{L}'$, introduced in Eq. (1), is the angular momentum of the system relative to its centre of mass and, as such, it is independent of the centre of mass motion. $\boldsymbol{L}'$ is variously referred to as intrinsic angular momentum, spin angular momentum, or simply *spin*. For a continuous body of volume $V$, $\boldsymbol{L}'$ can be expressed as:

$$\boldsymbol{L}' = \iiint_V \rho (\mathbf{r}' \times \mathbf{v}) \mathrm{d}V , \tag{3}$$

where $\rho$ denotes mass density, $\mathbf{r}'$ is the vector position of the generic point $P$ of the body relative to the centre of mass $C$, and $\mathbf{v}$ is the velocity vector of the same point. Clearly,

$$\mathbf{r}' = \mathbf{r} - \mathbf{c} , \tag{4}$$

where $\mathbf{r}$ is the position vector of point $P$ of the body with respect to point $O$ (Fig. 1).

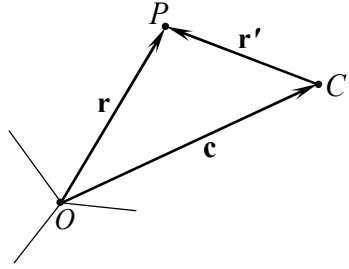

**Fig. 1.** Position vectors involved in definitions (2)–(4).

The angular momentum balance (Euler's 2$^{nd}$ law of motion) states that

$$\frac{d\boldsymbol{L}}{dt} = \boldsymbol{\Gamma},  \qquad (5)$$

where $\boldsymbol{\Gamma}$ is the torque of all the forces applied to the system or the considered part. Both $\boldsymbol{L}$ and $\boldsymbol{\Gamma}$ are understood to be relative to the same point $O$. If point $O$ is taken as coinciding with the centre of mass $C$ of the system, then $\mathbf{c} \equiv \mathbf{0}$ and from Eq. (2) we have that $\boldsymbol{L}_{\text{orb}} = \mathbf{0}$. From Eqs. (1) and (5) it then follows that

$$\frac{d\boldsymbol{L}'}{dt} = \boldsymbol{\Gamma}', \qquad (6)$$

where $\boldsymbol{\Gamma}'$ is the resultant torque relative to $C$ of all the forces and torques that act on the system. Eq. (6) is entirely general. It applies to any system or any part of it, irrespective of whether it is deformable or not and irrespective of the motion of its centre of mass, and thus of its acceleration. The quantities $\boldsymbol{L}'$ and $\boldsymbol{\Gamma}'$ are both defined in the inertial coordinate system considered to describe the motion, and therefore they do not depend on the acceleration that $C$ may undergo.

In particular, Eq. (6) applies to each vortex segment in a fluid flow. In this case, $\boldsymbol{\Gamma}'$ represents the total torque acting on the segment and $\boldsymbol{L}'$ is the segment's spin. However, if $\boldsymbol{\Gamma}'$ is always equal to zero during the flow, then $\boldsymbol{L}'$ cannot change and remains equal to zero if it was equal to zero initially. In other words, to acquire a spin and hence to become a vortex, the fluid making up a vortex segment has to be acted upon by a torque, at least for some time during the flow. No spinning vortex segment, and hence no turbulent flow can otherwise be produced.

Now observe that, the balance of linear momentum (Euler's 1$^{st}$ law of motion):

$$\iiint_V \rho \mathbf{a}\, dV = \iiint_V \rho \mathbf{f}\, dV + \iint_S \mathbf{t}^{(n)}\, dA, \qquad (7)$$

states that the resultant of all the forces (inclusive of the inertia forces, $-\iiint_V \rho \mathbf{a}\,\mathrm{d}V$) applied to any volume *V* of a fluid must vanish. In Eq. (7), ρ indicates mass density, **f** is the body force per unit mass, $\mathbf{t}^{(n)}$ is the stress vector, i.e. the force per unit area *A*, acting on the surface *S*, of the considered volume of fluid, while **a** = *d***v**/*dt* is the acceleration vector, i.e., the material time derivative of the velocity vector **v**. The vanishing of the resultant of the applied forces implies that their resultant moment vanishes relative to every point in space and, in particular, relative to the centre of mass of *V*. This entails that $\boldsymbol{\Gamma}' = \mathbf{0}$ unless there is a torque applied to the considered volume, thus making the resultant moment different than zero, even if the resultant of the applied forces vanishes.

Notice that, when applied to an infinitesimal volume element, Eq. (7) reduces to the local form of the linear momentum balance equation, also referred to as Cauchy's 1st law of motion:

$$\rho \mathbf{a} = \rho \mathbf{f} + \mathrm{div}\,\boldsymbol{\sigma} \qquad or \qquad \rho\, a_i = \rho\, f_i + \sigma_{ik,k}\,. \qquad (8)$$

The quantity **σ** introduced here denotes the stress tensor. It is related to $\mathbf{t}^{(n)}$ by the expressions:

$$\mathbf{t}^{(n)} = \boldsymbol{\sigma}\, \mathbf{n} \qquad or \qquad t_i^{(n)} = \sigma_{ij}\, n_j\,. \qquad (9)$$

Classical fluid mechanics of non-polar fluids does not include torques among the actions that may be applied to a viscous fluid. In these conditions, any rotation of any part of the fluid is strictly orbital. As a consequence, no spin and thus no turbulent vortex can be predicted by the theory. However, as shown in the next sections, vortex producing torques may arise in a fluid as the shear stress exceeds a certain limit, this is a phenomenon that is well within the scope of the classical (i.e., non-polar) theory of viscous fluid flow. The study of turbulence can thus be brought within the framework of classical mechanics.

For convenience later on, we record here the angular momentum balance equation (Euler's 2nd law of motion) in a slightly more general form than the one usually considered for a classical fluid:

$$\frac{d}{dt}\iiint_V \rho(\mathbf{l} + \mathbf{r}\times\mathbf{v})\,\mathrm{d}V = \iiint_V \mathbf{r}\times\rho\mathbf{f}\,\mathrm{d}V + \iint_S \mathbf{r}\times\mathbf{t}^{(n)}\,\mathrm{d}A\,, \qquad (10)$$

where the vector **l** represents the specific spin angular momentum per unit mass or specific spin for short. In the present form, Eq. (10) particularises a still more general angular momentum balance, which is often introduced when dealing with polar fluids and is widely discussed in the literature (see for example [13-15]). By using Eq. (8), the local form of Eq. (10) can be shown to reduce to (cf., e.g., Sect. 7.6 of [16]):

$$\rho\frac{dl_i}{dt} = \mathrm{e}_{ikj}\,\sigma_{jk}\,, \qquad (11)$$

where $\varepsilon_{ikj}$ is the permutation symbol.

In deriving Eq. (11), special attention should be paid to the sequence of the indices k and j appearing in tensor $\boldsymbol{\varepsilon}$. The correct sequence depends on how the stress tensor is defined in relation to $\mathbf{t}^{(n)}$. The present paper assumes that this relation is given by Eq. (9), which is consistent with what was done, for instance, in Truesdell *et al.* [12] and Leigh [16]. However, this is not the only choice adopted in the literature. Alternatively, the relation between $\boldsymbol{\sigma}$ and $\mathbf{t}^{(n)}$ is equally often assumed to be $\mathbf{t}^{(n)} = \mathbf{n}\boldsymbol{\sigma} = \boldsymbol{\sigma}^T \mathbf{n}$ or $t_i^{(n)} = \sigma_{ji} n_j$. In this case, the summation takes place over the first index of $\boldsymbol{\sigma}$ rather than the second, as in Eq. (9). If this alternative is adopted, the correct local form of the angular momentum balance can be obtained from Eq. (11) by exchanging the position of the indices k and j attached to tensor $\boldsymbol{\varepsilon}$ in that equation or, equivalently, by changing the sign of the right-hand side of the same equation.

It may also be observed that the right-hand side of Eq. (11) defines a vector $\boldsymbol{\gamma}$ of components

$$\gamma_i = e_{lkj}\, \sigma_{jk}. \tag{12}$$

This vector is the opposite of twice the axial vector usually associated with the antisymmetric part of a second-order tensor such as $\boldsymbol{\sigma}$ (cf., e.g., Sect. 2.14 of [16]). It has the dimensions of torque divided by volume and represents the *specific torque* per unit volume that would result from the antisymmetric part of the stress tensor. More explicitly, Eq. (12) can be written as

$$\gamma_1 = e_{1kj}\, \sigma_{jk} = \sigma_{32} - \sigma_{23}, \tag{13}$$

$$\gamma_2 = e_{2kj}\, \sigma_{jk} = \sigma_{13} - \sigma_{31}, \tag{14}$$

and

$$\gamma_3 = e_{3kj}\, \sigma_{jk} = \sigma_{21} - \sigma_{12}, \tag{15}$$

which provide a handy means to calculate the components of the specific torque once the stress tensor $\boldsymbol{\sigma}$ is known.

By means of $\boldsymbol{\gamma}$, Eq. (11) can be expressed in vector form as:

$$\rho \frac{d\mathbf{l}}{dt} = \boldsymbol{\gamma}. \tag{16}$$

From Eqs. (13)–(15), it is immediate to verify that $\boldsymbol{\gamma} \equiv \mathbf{0}$ if $\boldsymbol{\sigma}$ is symmetric. The importance of Eq. (16) should not be underestimated. It shows that a change in the spin angular momentum of the fluid can only be produced at the points where the stress tensor $\boldsymbol{\sigma}$ happens to be non-symmetric.

The classical approach to viscous fluids ignores the spin angular momentum of the fluid, and accordingly excludes vector $\mathbf{l}$ from Eq. (10). In that approach, it is then inferred that the left-hand side of Eq. (11) should vanish, and that therefore the stress tensor should be symmetric. This conclusion is strictly valid as long as no spin is produced, i.e., as long as no turbulent vortex is generated in the fluid. As discussed in the next section, the stress tensor of a viscous fluid ceases to be symmetric if the velocity gradient exceeds a definite limit. When this happens, the antisymmetric part of the stress tensor generates spin angular momentum in the fluid, according to Eqs. (10), (11), or (16).

### III. LIMIT TO STRESS TENSOR SYMMETRY

Fluid viscosity generates friction forces at the interface of adjacent layers of fluid that slide over one another due to their different velocities. Fig. 2(a) provides a sketch of the phenomenon for two layers of infinitesimal thickness $dx_2$, sliding over their contact surface while proceeding in a plane rectilinear flow in the $x_1$ direction. The friction forces $\mathbf{t}_f$, that the two layers exchange per unit area across the contact surface are opposite to each other. The value of $\mathbf{t}_f$ depends on the relative velocity of the layers or, better, on the velocity gradient in the normal direction to the sliding surface. In the case of a linear viscous fluid, this dependence is linear and the modulus of $\mathbf{t}_f$ is given by:

$$t_f = \eta \frac{dv_1}{dx_2}, \tag{17}$$

where the constant $\eta$ is the viscosity coefficient of the fluid.

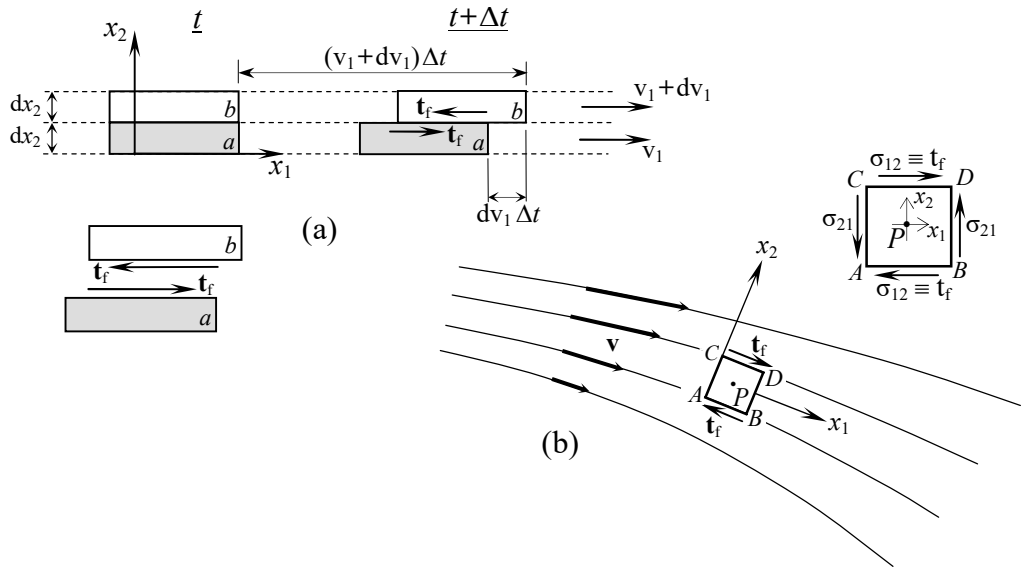

**Fig. 2.** (a) Viscous friction forces $\mathbf{t}_f$ at the contact surface between two plane layers of fluid flowing at different velocities. (b) Stress components on a cubic element about point $P$ of a non-uniform streamline flow.

This phenomenon takes place in every non-uniform viscous flow. Let us refer, for simplicity, to a streamlined (i.e., non-turbulent) flow, such as the one represented in Fig. 2. For simplicity, we ignore any curvature effect (which is strictly valid in the limit of a plane flow). At any given point $P$ of the fluid, let us consider a cubic volume element $ABCD$, with two faces, $AC$ and $BD$, normal to the streamline at $P$. We refer this element to the triad of axes $(x_1, x_2, x_3)$ specified in the figure (the $x_3$-axis being perpendicular to the plane of the figure and directed outward). The shear stress component on faces $AB$ and $CD$ of the element is due to the friction with the adjacent layers of fluid and coincides with the friction force $\mathbf{t}_f$ introduced above. Therefore, by denoting $\boldsymbol{\sigma}$ the stress tensor at $P$, we have that, in the considered system of reference,

$$\sigma_{12} = t_f, \tag{18}$$

where $\sigma_{12}$ denotes the component of $\boldsymbol{\sigma}$ acting on the face normal to $x_2$ and directed along the $x_1$-axis.

No friction acts on faces $AC$ and $BD$ of the cube. The reason is that the fluid velocity parallel to these faces is zero (the faces are normal to the streamlines). We should conclude, therefore, that the shear stress $\sigma_{21}$ on these faces should vanish. That is

$$\sigma_{21} = 0. \tag{19}$$

When considered together with Eq. (18), however, this equation is at odds with the stress symmetry relation

$$\sigma_{ij} = \sigma_{ji}, \tag{20}$$

which in the present case requires that

$$\sigma_{21} = \sigma_{12} = t_f. \tag{21}$$

The symmetry of the stress tensor is often introduced directly through the constitutive equations of the fluid, and the angular momentum balance is satisfied trivially in the absence of torques per unit volume. As no friction acts on faces $AC$ and $BD$, the non-vanishing value of $\sigma_{21}$ requested by the symmetry relation (20) must be the consequence of the deformation of the cube.

Now, there is no physical limit to the maximum value of the friction force that can be applied on a fluid surface, provided that the sliding velocity over the same surface is appropriately large. We must admit, therefore, that the value of $\sigma_{12}$ on the cube of Fig. 2(b) can be made as large as we wish. However, there is a limit to the shear stress resistance of a fluid. This is due to the finite capacity of any real material to oppose deformation, before breaking, yielding, or somehow losing the capacity to react. On faces $AC$ and $BD$, therefore, the fluid element in Fig. 2(b) can exert the stress $\sigma_{21}$ requested by Eq. (21), only if the value of $t_f$ is appropriately small.

Let $\tau_y$ denote the *ultimate shear stress* of the fluid. This is the value of the shear stress resistance of the fluid under pure shear stress. As discussed in Paglietti [17] and [18], $\tau_y$ can be determined from experiments on plane Couette flow. As observed above, the value of the stress component $\sigma_{12}$ applied to faces $AB$ and $CD$ of the considered cube is due to viscous friction and it can be made as large as we wish, since Eq. (18) is valid irrespective of the value of $t_f$. However, if $t_f$ exceeds the limitation

$$t_f \leq \tau_y, \tag{22}$$

or, equivalently given Eq. (17), if

$$\frac{dv_1}{dx_2} > \frac{\tau_y}{\eta}, \tag{23}$$

then, as apparent from Eq. (18), the applied stress $\sigma_{12}$ becomes larger than the maximum value $\tau_y$ of the stress component $\sigma_{21}$ that the fluid exert. This means that Eq. (21)$_1$ cannot be met and the stress tensor ceases to be symmetric. Under these conditions the off-diagonal components of the stress tensor result in a torque applied to the considered element of fluid. The specific value of this torque, per unit volume, is $\gamma$ and can be calculated from Eqs. (12)–(15).

A precise determination of the value that the stress component $\sigma_{21}$ assumes once $\sigma_{12}$ exceeds $\tau_y$ is outside the realm of today's experimental knowledge. We can however observe that in most fluids the value of $\tau_y$ is comparatively small. For instance, as discussed in Paglietti [17], the value of $\tau_y$ for water at room temperature lies in the range $11 \cdot 10^{-3}$ Pa to $30 \cdot 10^{-3}$ Pa. In view of this, we shall assume that $\sigma_{21}$ drops to zero as $\sigma_{12}$ exceeds the limit $\tau_y$. Accordingly, the stress tensor at the points of the fluid where condition (23) applies will approximately be given by:

$$\sigma = \begin{vmatrix} -p & t_f & 0 \\ 0 & -p & 0 \\ 0 & 0 & -p \end{vmatrix} \quad (if \ t_f > \tau_y), \tag{24}$$

where $p$ denotes pressure. Clearly, the greater the value $\sigma_{12}$ is in excess of $\tau_y$, the better this assumption.

Of course, if $\sigma_{12}$ does not exceed $\tau_y$, the stress tensor components are:

$$\sigma = \begin{vmatrix} -p & t_f & 0 \\ t_f & -p & 0 \\ 0 & 0 & -p \end{vmatrix} \quad (if \ t_f \leq \tau_y). \tag{25}$$

In this case, the tensor $\sigma$ is symmetric, as assumed in the classical theory of viscous fluids. We can conclude, therefore, that the existence of a limit to the shear stress resistance of a viscous fluid produces a non-symmetric stress tensor if this limit is exceeded.

In the present case, the components of the torque $\gamma$ at the points of the fluid where the stress tensor ceases to be symmetric can easily be obtained by inserting the components of $\sigma$ given by Eq. (24) into Eqs. (13)–(15). We thus obtain:

$$\gamma_1 = 0, \quad \gamma_2 = 0 \quad and \quad \gamma_3 = -\sigma_{12} = -t_f = -\eta \frac{dv_1}{dx_2}. \tag{26}$$

The last equality in the expression of $\gamma_3$ follows from Eq. (17). The result (26) can be checked directly from Fig. 2(b), once we observe that the force acting on faces *AB* and *CD* is $t_f\, dx_1 dx_3$, while the distance between them is $dx_2$. In the considered case, $\sigma_{21} = 0$. The negative sign of $\gamma_3$ is because the torque rotates clockwise, and thus points in the negative direction of the $x_3$-axis, the considered coordinate system being right-handed.

## IV. NEAR-WALL VORTEX FORMATION ZONE

Let $\Sigma$ be any axially symmetric volume portion of a fluid. To be specific, we may visualise this portion as a circular cylinder of finite length. In this case, the cylinder axis $\xi$, coincides with the axial symmetry axis of $\Sigma$. Being a symmetry axis, $\xi$ is barycentric. From Eq. (6) it then follows that a torque $\boldsymbol{\Gamma'}$ parallel to $\xi$ has to be applied to $\Sigma$ to make it spin about its axis or to change the spin magnitude about the same axis. As observed in Sect. **II**, any different system of forces, not equivalent to the above torque, cannot affect the spin of $\Sigma$ about $\xi$.

Vortex segments are fluid portions, like $\Sigma$, that spin about their axial-symmetry axis. A vortex segment is set into motion by the torque which results from the shear stress at its surface and is due to the viscous friction between the adjacent layers of fluid flowing at different velocities. Viscous friction also contributes to damp a vortex once it is created, although it need not be the only or the leading cause of vortex damping.

To analyse the formation of a vortex segment, we focus on the flow near a wall. For simplicity, we refer to a steady-state flow near a plane wall, parallel to the flow direction. We also assume that the fluid is sufficiently far from other walls so that we can ignore their presence. In this case, the streamwise velocity of the fluid only depends on the distance from the considered wall. In the notation of Fig. 3, we have, therefore, that $v_1 = v_1(x_2)$. In the turbulent regime, the values of $v_1$ fluctuate over time. Thus, when referring to turbulent flow, we shall henceforth understand that $v_1(x_2)$ is the time average of $v_1$ at the distance $x_2$ from the wall. Unless otherwise stated, the same convention is understood to also apply to $\boldsymbol{\sigma}$, $\boldsymbol{\gamma}$, or, more generally, to any other quantity whose value may fluctuate due to flow turbulence.

In every turbulent flow over a smooth wall, there is always a thin layer adjacent to the wall, where the average streamwise velocity of the fluid varies linearly with the distance from the wall, starting from zero at the wall. This layer is referred to as the *viscous sublayer* or the *laminar sublayer*, although the fluid velocity in this layer is known to fluctuate in a seemingly random way both in intensity and direction ([19-21]). Denoting the shear wall stress as $t_w$, the slope of the velocity diagram at the wall is

$$\frac{dv_1}{dx_2} = \frac{t_w}{\eta}, \qquad (27)$$

as immediately follows from Eq. (17). This is also the slope of the entire viscous sublayer because $v_1$ varies linearly within it. Therefore, throughout the viscous sublayer the shear stress in the fluid is constant and equal to $t_w$.

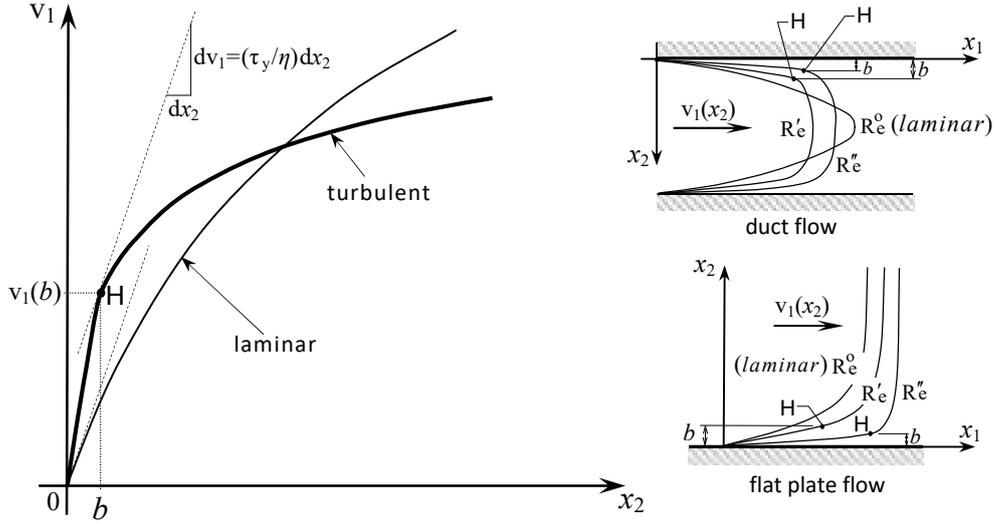

**Fig. 3.** Typical curves showing the fluid velocity $v_1$ as a function of the distance $x_2$ from the wall. The insets to the right illustrate the notation adopted. In the turbulent flow case, the shear stress of the fluid exceeds the limit stress $\tau_y$ in the whole interval $0 \leq x_2 < b$. Point H, where the slope of the curve equals the value $\tau_y/\eta$, determines the amplitude $b$ of this interval. No point H exists in the laminar case, as the slope to the curve is, in this case, consistently below the value $\tau_y/\eta$.

The thickness of the viscous sublayer is denoted as $b$ in Fig. 3. From the experiments we know that:

$$5y^+ \leq b \leq 7y^+ \qquad (28)$$

(cf., e.g., Panton [2]). Here, $y^+$ denotes the *viscous length*, defined as:

$$y^+ = \frac{\nu}{u*}, \qquad (29)$$

where $\nu = \eta/\rho$ is the kinematic viscosity, $\rho$ is mass density, and

$$u* = \sqrt{\frac{\tau_w}{\rho}}, \qquad (30)$$

is the *friction velocity*. In most practical applications, the value of $b$ for air and water varies between approximately 0.1 mm and 1 mm [20].

The above equations show that as the flow velocity increases or, more generally, as the Reynolds number associated with the flow increases, the slope of the velocity diagram of the viscous sublayer becomes steeper and *b* decreases. These features are well-known and supported by a large body of experimental evidence (cf., e.g., [4, 22-24] to quote just a few works on the argument). The same features are also apparent from the diagrams of Fig. 3, once it has been noted that the Reynolds numbers considered here are chosen in such a way that $\overset{o}{R}_e < \overset{'}{R}_e < \overset{''}{R}_e$, with $\overset{o}{R}_e$ in the laminar regime, and $\overset{'}{R}_e$ and $\overset{''}{R}_e$ in the turbulent range. We should observe, however, that the present analysis also applies to the case in which the slope of the velocity diagram at the wall remains constant at its laminar limit value, as the Reynolds number increases in the turbulent range.

As discussed in Paglietti [17] and [18], the shear stress of a laminar flow can never exceed the value $\tau_y$. This means, in particular, that the slope of the velocity curve of a laminar flow must meet the limitation

$$\frac{dv_1}{dx_2} \leq \frac{\tau_y}{\eta} \tag{31}$$

at every point of the curve. This limitation does not apply to turbulent flow. Thus, as previously observed, there is no *a priori* upper bound to the slope of the viscous sublayer, provided that the flow velocity, or better the flow's Reynolds number, can be increased appropriately.

At the points where inequality (31) is not met, inequality (23) applies. At these points, the stress ceases to be symmetric, as discussed in Sect. **III**. This results in a torque **γ** per unit volume, which generates spin angular momentum in the fluid, according to Eq. (16). Therefore, the regions of the fluid where limitation (31) is exceeded are the regions where turbulent vortices are created.

In Fig. 3, the letter H indicates the points where the slope $dv_1/dx_2$ of the velocity curve equals $\tau_y/\eta$. Since the slope of these curves decreases with the distance from the walls, the H points also mark the distance from the wall up to which a torque **γ** per unit volume can be formed in the fluid due to excessive shear stress. In practice, this distance coincides with the thickness *b* of the viscous sublayer, as the curvature at point H is usually quite pronounced. Beyond point H, relation (31) is met, and **γ** cannot be produced. For the considered flows, therefore, all vortex production can only take place in the fluid near the wall, up to distance *b*.

Relation (31) has profound implications for the velocity field of a viscous fluid. When relation (31) is met, the fluid can react with a symmetric stress tensor to equilibrate the shear stress arising from the viscous friction with the adjacent fluid. In this case, the flow is entirely regulated by the linear momentum balance because, as observed in Sect. **III**, the mere symmetry of the stress tensor suffices to meet the angular momentum balance.

On the other hand, if limitation (31) is exceeded, which happens if the flow's Reynolds number is appropriately high, then the shear stress resistance of the fluid is exceeded. In this case, the stress tensor ceases to be symmetric, thus resulting in a non-vanishing specific torque γ at the points where limitation (31) is violated. The part of the fluid where γ ≠ **0** is set into rotation (spin) about its centre of mass and acquires a relative motion with respect to the surrounding fluid. As dictated by Eq. (16), the rate of spinning of this part of fluid increases until the opposing torque resulting from the friction with the surrounding fluid does not equilibrate the driving torque resulting from γ. A vortex is thus produced.

V.       THE VORTICAL STRUCTURE OF THE VISCOUS SUBLAYER

By introducing two different ways in which the fluid responds to the applied forces, limitation (22) or, equivalently, limitation (31) is likely to produce discontinuities in the stress field, and thus in the velocity field of the fluid. This occurs, in particular, at the interface between the spinning vortices and the surrounding fluid. These discontinuities greatly increase the analytical complexity of the solution to the fluid dynamic problem. This is especially true when the parts of the fluid where limitation (31) is exceeded are small, numerous, and in relative motion to each other, as happens in a turbulent flow. Under these conditions, a rigorous analytical solution of the equations of motion is impracticable. The simple, though approximate, analysis of vortex production and diffusion presented in this section and the following one may facilitate a better comprehension of the phenomenon.

To begin with, let us consider a rigid circular cylinder rolling without slipping over a plane. The latter is taken as coinciding with the ($x_1$, $x_3$)-plane. The cylinder axis is assumed to be parallel to the $x_3$-axis. A cross-section of the cylinder, normal to the cylinder axis, is represented in Fig. 4. The $x_1$-axis coincides with the trace of the plane of rolling. The cylinder's angular velocity is indicated as $\dot{\theta}$ and is represented by a vector parallel to the cylinder axis. The nonslip condition means that the axis of instantaneous rotation coincides with the contact line between the cylinder and the plane of rolling. The centre of instantaneous rotation of the considered cross-section is, therefore, its contact point, $O'$, with the $x_1$-axis. Thus, the velocity vector **v** of any point $P$ of that section is directed as the normal to the line $PO'$, and its magnitude is given by

$$v = \dot{\theta}\, d , \qquad (32)$$

where d is the length of segment $PO'$. If β is the angle that $PO'$ forms with the vertical, the component of **v** along the $x_1$-axis is

$$v_1 = \dot{\theta}\, d \cos\beta = \dot{\theta}\, x_2 , \qquad (33)$$

where $x_2$ is the distance of $P$ from the said axis. This equation shows that the horizontal component of the velocity of the points of the considered cylinder is a linear function of their distance from the plane.

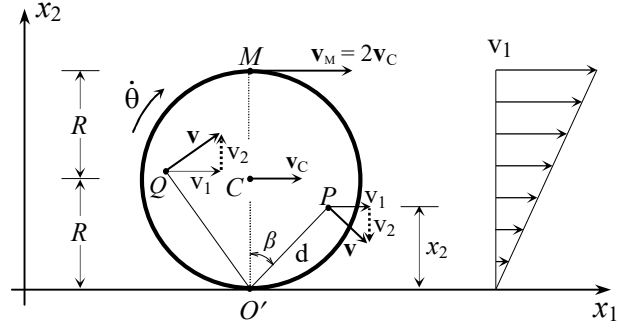

**Fig. 4.** Cross-section of a rigid cylinder rolling without slipping over a horizontal plane. The diagram to the right shows that the horizontal component $v_1$ of the velocity of the points of the cylinder vanishes at the contact point $O'$, and increases linearly with the distance $x_2$ from the plane.

Let us now consider a circular cylindrical portion of fluid belonging to the viscous sublayer of a steady-state fluid flow over a plane wall. We shall refer to this portion as the cylinder $\Sigma$ or $\Sigma$ for short. As before, we assume that the wall coincides with the plane $(x_1, x_3)$ and that the cylinder axis is parallel to the $x_3$-axis. The diameter of $\Sigma$ is supposed to equal the thickness $b$ of the viscous sublayer. The generatrix of the cylinder $\Sigma$ tangent to the wall intersects the $x_1$-axis at a point $O'$, as in Fig. 4. Because $\Sigma$ belongs to the viscous sublayer, the streamwise component of the velocity of its points is a homogeneous linear function of $x_2$. Its expression is, therefore,

$$v_1 = v_1(x_2) = \frac{t_w}{\eta} x_2, \tag{34}$$

as immediately follows from Eq. (27) and from the nonslip condition $v_1(0) = 0$. Remember that Eq. (34) applies to the average value of $v_1$ since the flow is turbulent. Of course, when dealing with the viscous sublayer, we must understand that $t_w > \tau_y$ because the viscous sublayer only forms in the turbulent regime. From Eq. (34) we calculate, in particular, that the fluid velocity at the point of $\Sigma$ at distance $b$ from the wall (i.e., on the upper surface of the viscous sublayer) is given by:

$$v_1(b) = \frac{t_w}{\eta} b. \tag{35}$$

As discussed in Sect. **III**, the laminar flow is not compatible with Eq. (34) and with the angular momentum balance equation, if $t_w > \tau_y$. Therefore, the flow at the points of the considered cylinder cannot be laminar even though the streamwise component of the velocity of its points varies linearly. The experiments have widely confirmed both the linearity of $v_1$ and the turbulence of the viscous sublayer.

The laminar regime, however, is not the only possibility for Eq. (34) to be valid. A glance at Eq.

(33) shows that if the cylinder Σ rolls rigidly on the wall, without slipping, at the angular velocity

$$\dot{\theta} = \frac{v_1(b)}{b} = \frac{t_w}{\eta}, \qquad (36)$$

then the velocity component $v_1$ of its points has exactly the expression (34). Under these conditions, the streamwise component $v_1$ of the particles of fluid that make up the cylinder is a homogeneous linear function of the distance of the particles from the wall, as prescribed by Eq. (34), and yet the flow is far from laminar.

In view of the above observations, the fact that in the viscous sublayer the flow is not laminar while the velocity component $v_1$ varies linearly with the distance from the wall suggests that the same sublayer is made of a sequence of cylinders Σ that roll rigidly without slipping on the wall, see Fig. 5(a). The cylinders proceed at a certain interaxis λ from each other. They must all roll at the same angular velocity to make the $v_1$ diagram of their points consistent with that of the viscous sublayer.

As for the fluid of the viscous sublayer that is located in the interspaces between the cylinders Σ, we observe that, in steady-state conditions, all the Σ cylinders roll without slipping at the same angular velocity. Since the cylinders are in rigid-body motion, it follows that the average value of the $v_1$ component of the velocity of the portion of fluid between two subsequent cylinders must be given by Eq. (34), no matter how complex the instantaneous velocity of the points of this portion. In this way, the viscous sublayer can exhibit the linear average streamwise velocity field $v_1$ specified by Eq. (34), even though its motion is far from laminar. This conclusion agrees with the well-established experimental evidence concerning both the non-laminar character of the flow of the viscous sublayer and the linear dependence of the average value $v_1$ of its points from the distance from the wall.

The driving torque of each cylinder Σ of the sequence of cylinders described above comes from the viscous friction, $t_f$, acting on plane $\partial B$ at distance $b$ from the wall, Fig. 5(a). This friction produces a force directed as the $x_1$-axis and applied to the point of the cylinder surface located at distance $b$ from the wall (cf. the analogous point $M$ of Fig. 4). Without introducing undue restriction to the present analysis, the value of this force per unit height of cylinder (i.e., per unit length in the direction of the $x_3$-axis) can be expressed as $a\, t_f$, where $a$ defines an appropriate length across the cylinder, as indicated in Fig. 5 (remember that $t_f$ is a force per unit area). The moment of this force relative to the wall gives the driving torque $m_f$ per unit height of cylinder. Its amplitude is, therefore:

$$m_f = a\, t_f\, b = t_w\, a\, b, \qquad (37)$$

where we exploited the fact that in the viscous sublayer $t_f = t_w$, due to the linearity of $v_1 = v_1(x_1)$. In steady-state conditions, the torque $m_f$ is balanced by the torque produced by the friction forces acting on the cylinder surface at the interfaces with the adjacent fluid of the considered sublayer. From simple symmetry considerations, we infer that half of $m_f$ is equilibrated at the downstream interface and the other half at the upstream interface, Fig. 5(b).

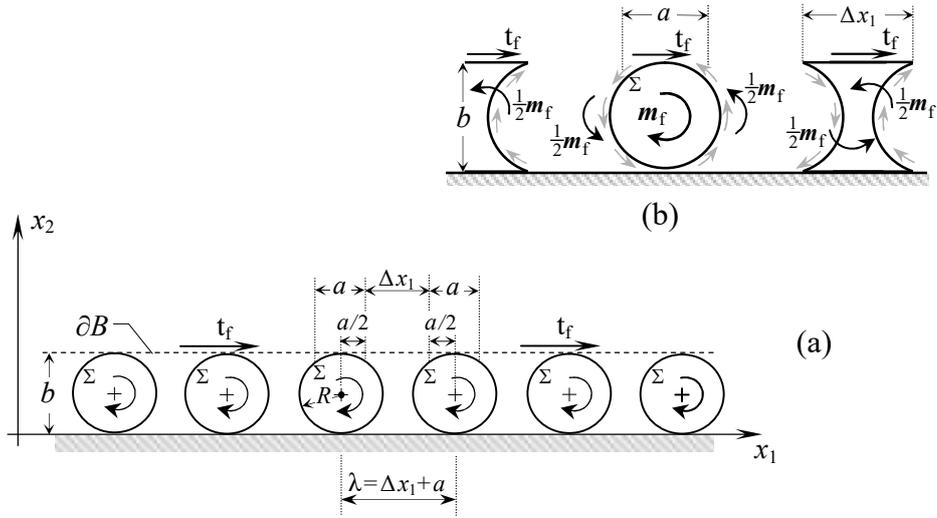

**Fig. 5.** (a) Sequence of fluid cylinders Σ in the viscous sublayer. The cylinders roll at same angular velocity without slipping on the wall. (b) Driving torque $m_f$ on a cylinder Σ, and torques ½ $m_f$ produced by the viscous friction forces (grey arrows) at the interfaces between the rolling cylinder and the adjacent fluid of the viscous sublayer.

The segments of the viscous sublayer that are confined by two successive cylinders, Fig. 5(b), are acted upon by the torque 2 (1/2 $m_f$) = $m_f$, applied to them by the adjacent cylinders. In steady-state conditions, this torque must balance the torque $m'_f$ produced by the viscous friction $t_f$ acting on the segment surface on plane $\partial B$. The value of this torque, per unit length in the $x_3$ direction, is:

$$m'_f = \Delta x_1 \, t_f \, b = t_w \, \Delta x_1 \, b, \tag{38}$$

where $\Delta x_1$ is the length specified in the above figure. By equating Eq. (37) to Eq. (38) it is immediate to conclude that

$$\Delta x_1 = a. \tag{39}$$

As illustrated in Fig. 5(a), the interaxis between two successive cylinders Σ is given by

$$\lambda = \Delta x_1 + a = 2a, \tag{40}$$

the last equality following from Eq. (39). Of course,

$$\lambda \geq b, \tag{41}$$

because the cylinders cannot superimpose each other. Obviously, $a$ cannot exceed $b$. From Eqs. (40)$_2$ and (39) it then follows that

$$b \leq \lambda \leq 2b, \tag{42}$$

which shows that the thickness of the viscous sublayer sets both an upper and a lower bound to the interaxis $\lambda$ of the cylinders Σ.

The rolling cylinder structure of the viscous sublayer considered here, requires that the component of the fluid particle velocity normal to the wall (i.e., $v_2$) should fluctuate significantly, even at a very short distance from the wall. The dashed arrows in Fig. 4 that represent the $v_2$ velocity components at two points of the rotating cylinder illustrate this property schematically. This phenomenon is sometimes referred to as "wall penetration". It has been known for a long time experimentally (cf., e.g., [25, 26] and the references quoted within) and it provides further evidence of the validity of the proposed approach.

## VI.  VORTEX MIGRATION

All turbulent vortices originate in the viscous sublayer, as this is the only region of the flow where torques are applied to the fluid. Once a cylinder $\Sigma$ forms in this sublayer, it is lifted, veered, and broken into parts under the action of several kinds of forces, the most intriguing of which depend on the cylinder's spin. The vortex segments migrate and diffuse through the flowing fluid, which thus becomes turbulent.

As long as they spin rigidly about their axis, the cylinders $\Sigma$ or their segments behave as rigid inclusions in the flowing fluid. These inclusions are free from their own weight because they are made of the same material as the surrounding fluid. This enhances their mobility under the slightest action that may happen to be applied to them. All the dissipation of their kinetic energy is confined to the viscous friction on their surface. No dissipation occurs in the fluid within them as long as they move and rotate rigidly. As a result, the dissipation per unit volume of a spinning vortex segment is likely to be quite low if compared with the average specific dissipation of the flow. This would explain why these vortices, once they detach from the wall, persist in their motion for thousands of wall units downstream while traveling through the flow [8].

To better appreciate the physics of turbulent vortex production, we shall frequently refer to the particular example of a flow with a viscous sublayer of thickness $b = 0.5$ mm $= 0.5 \; 10^{-3}$ m and streamwise velocity at distance $b$ from the wall equal to $v_M = v_1(b) = 1$ m/sec. In this case, according to the analysis of the previous section, the cylinders $\Sigma$ possess a radius $R = b/2 = 0.025 \; 10^{-3}$ m, a translational velocity $v_C = v_1(b/2) = 0.5$ m/sec, and according to Eq. (36), an angular velocity $\dot{\theta} = 2 \; 10^3$ rad/sec = 12,570 rpm. We assume that the fluid is water at 20°C, which means a mass density of $\rho = 10^3$ Kg/m$^3$, and a viscosity coefficient of $\eta = 10^{-3}$ N sec/m$^2$.

The forces acting on the cylinders $\Sigma$ and their effects are discussed below.

### A.  Lifting force

Due to their reduced diameter, the cylinders $\Sigma$ must roll at a high rotation rate $\dot{\theta}$ to keep pace with the flowing fluid without slipping on the wall, in agreement with the value of $\dot\theta$ calculated in the example above. The high angular velocity produces a consistent Magnus effect on the rolling cylinders, resulting in a lifting force $F_L$, per unit length of the cylinder axis, whose general expression is:

$$F_L = \tfrac{1}{2} c_L \, \rho \, U_\infty^2 \, D, \qquad (43)$$

see for example references [27-30]. In this equation, $c_L$ is the lift coefficient, $U_\infty$ is the oncoming flow velocity, and $D$ is the cylinder diameter (coinciding with $b$ in the case of the cylinders $\Sigma$). Since in the present case the velocity of the oncoming flow is not uniform, we shall approximate $U_\infty$ with the average streamwise velocity $v_C = v_1(b/2)$ of the fluid of the viscous sublayer. The lifting force acting on $\Sigma$ will accordingly be evaluated approximately as

$$F_L = \tfrac{1}{2} c_L \, \rho \, v_C^2 \, b. \qquad (44)$$

It should be noted, however, that for the kinds of flow considered in this paper the average velocity of the fluid impinging the cylinder increases as the cylinder moves away from the wall.

The lift coefficient $c_L$ depends both on the Reynolds number

$$\mathrm{Re} = \frac{\rho U_\infty D}{\eta} \qquad (45)$$

and the rotation rate of the cylinder through the dimensionless parameter

$$\alpha = \frac{D \dot\theta}{2 \, U_\infty}. \qquad (46)$$

In the particular example considered above, by assuming $U_\infty = v_C = 0.5$ m/sec we calculate Re = 250 and $\alpha = 1$. As remarked in Stojković *et al*. [29], for $\alpha < 2$ the coefficient $c_L$ is almost independent of Re and varies linearly with $\alpha$. There is evidence, moreover, that this coefficient also depends on the distance of the cylinder from the wall [31, 32]. However, the experimental data on $c_L$ are scarce and do not allow for a precise assessment of its value. For this reason, based on the data reported in the literature, we shall conservatively assume that $c_L = 1$. Therefore, in the considered example the lifting force per axial metre of $\Sigma$ can be estimated from Eq. (44) and it has the value

$$F_L = 0.00625 \text{ N/m}. \qquad (47)$$

To get an idea of the relative magnitude of this force, we may observe that it is about three times the weight per metre cylinder $\Sigma$, which amounts to 0.002 N/m, as can easily be calculated from the data given above.

The lifting effect considered here takes the $\Sigma$ cylinders away from the viscous sublayer. Therefore, new $\Sigma$ cylinders are continuously produced in the same sublayer to replace the lost ones and re-establish a dynamic equilibrium. The phenomenon absorbs energy from the force that drives the flow because it subtracts kinetic energy from the fluid near the wall, since the spinning vortices move from the wall toward the main flow. This energy is dissipated into heat far from the wall, as the vortices fade away due to the viscous friction with the surrounding fluid. The higher the flow velocity, the stronger the lifting force, the greater the number of Σ cylinders that leave the viscous sublayer per unit time, and the larger the rate of energy dissipation.

### B. Vortex veering by gyroscopic precession

In their real proportions, the cylinders Σ are thin cylindrical filaments with a very small diameter over length ratio. These filaments are easily broken into parts because their tiny cross-section can only oppose small internal actions to the external disturbances and because a fluid material has little or no resistance to tensile stress. As a consequence, each filament will soon reduce to a more or less precise alignment of separate Σ cylinder segments, spinning about their axis with a comparatively large angular momentum $L_s$ per unit length. The large value of $L_s$ puts these segments at the mercy of the smallest disturbances that may change their orientation dramatically through gyroscopic precession, as discussed below.

As in any rigid body, any disturbance that modifies the rigid-body motion of a vortex segment can be thought to arise from a force acting on the segment centre of mass and a torque. The force affects the motion of the centre of mass of the segment; the torque affects its spin. In Fig. 6, the torque is denoted as $\mathbf{\Gamma}_F$ and is also visualised by a couple of two opposite forces $\mathbf{F}$ acting on the vortex segment on a plane normal to $\mathbf{\Gamma}_F$ passing through the segment axis. At their initial formation, all vortex segments are oriented spanwise with their axis parallel to the $x_3$-axis. Because of their spin, the application of a non-axial torque produces a gyroscopic precession of the segment, which makes it rotate about a precession axis, $k$-$k$, normal to both $\mathbf{\Gamma}_F$ and $\mathbf{L}_s$ (see Fig. 6).

Figure 6(a) represents the case of a spanwise vortex segment whose axis is initially parallel to the $x_3$-axis. The segment is acted upon by a disturbing torque $\mathbf{\Gamma}_F$ parallel to the $x_2$-axis. The disturbance initiates a rotational motion of the segment about the precession axis $k$-$k$, which lasts as long as the disturbing torque is applied. In a gyroscopic precession, the direction of the rotation is always the one that brings $\mathbf{L}_s$ towards $\mathbf{\Gamma}_F$. In the present case, this is a clockwise rotation. The angular velocity of precession is proportional to $\mathbf{\Gamma}_F$, as follows from to the angular momentum balance (6). The example in Fig. 6(a) shows how the considered disturbance can, in particular, transform the original vortex segment into a transverse vortex that is normal to the wall, assumed to coincide with the ($x_1$, $x_3$)-plane. In this case, the precession angle is $\pi/2$ and the spin of the cylinder, in its final position, is anticlockwise. The opposite disturbance, $-\mathbf{\Gamma}_F$, would produce the opposite precession angle, $-\pi/2$, and a final spinning in the clockwise direction. This may explain the formation of two nearby parallel vortices of opposite spin sometimes observed in the experiments. Of course, depending upon the intensity of $\mathbf{\Gamma}_F$ and its duration, any other angle of precession can be produced. For more details on the phenomenon of gyroscopic precession, the reader may refer to Kleppner *et al.* [11].

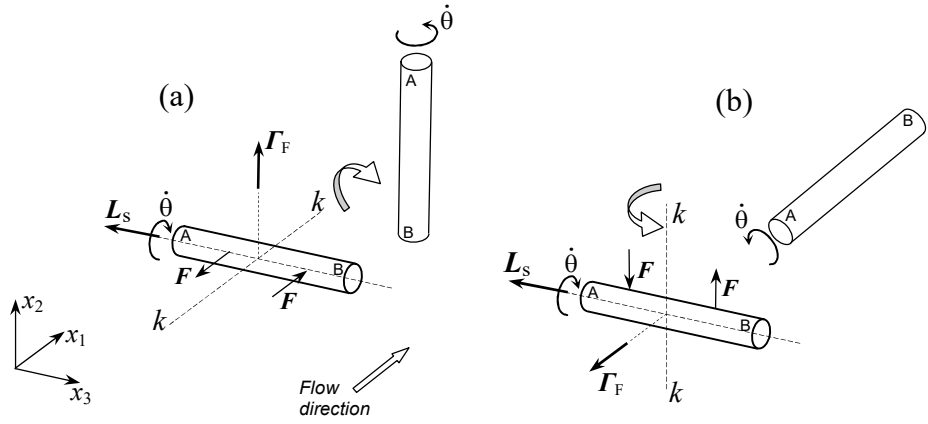

**Fig. 6.** Gyroscopic precession of a spinning vortex segment. (a) The disturbing torque, $\Gamma_F$, is parallel to the $x_2$-axis and turns into a vortex segment normal to the wall, a spanwise vortex segment which is initially parallel to the wall. (b) The same spanwise vortex segment is turned into a streamwise vortex segment by a disturbing torque $\Gamma_F$ parallel to the $x_1$-axis. In both cases, the axis of precession is indicated as $k$-$k$. The wall coincides with the $(x_1, x_2)$-plane.

On the other hand, as shown in Fig. 6(b), a disturbing torque parallel to the $x_3$-axis can transform the same spanwise vortex segment initially considered in Fig. 6(a) into a streamwise vortex segment. Again, the spin of the resulting vortex will be anticlockwise or clockwise, depending on the sign of the disturbance applied to the initial vortex, and hence on the sign of the precession angle produced. The case represented in Fig. 6(b) refers to a final anticlockwise spin.

Although all vortex segments are born spanwise, they are lifted from the wall via the Magnus effect and reoriented in different directions by small disturbances that result in large precession angles. Both phenomena are strictly dependent on the axial spin of the vortex segment on which they act. They apply to the myriad of vortex segments that are created in the viscous sublayer. As a result, a large portion of the fluid above the wall becomes teemed with a chaotic multitude of turbulent vortices with different orientation and even opposite spins, the origin of which is, at first, quite baffling [4, 33, 34]. The situation is further complicated by the viscous interaction between the spinning segments and the surrounding fluid. This interaction blurs the velocity field at the interface between the vortex segments and the surrounding fluid, thus masking the original dimensions of the segments and the way in which they group and interact together.

## VII.   CONCLUSIONS

Each vortex of the myriad of little vortices that populate a turbulent flow needs a torque to be created.

This is a consequence of the angular momentum balance law. In viscous fluids, the vortex producing torques are generated when the viscous friction exceeds the shear stress resistance of the fluid. When this happens, the stress tensor ceases to be symmetric and applies a torque per unit mass to the fluid. The involved portion of fluid starts to spin about a barycentric axis and generates a turbulent vortex as a result.

In wall-bounded turbulent flows of viscous fluids, the region where the velocity gradient, and thus the viscous friction, reaches the highest value is a tiny region adjacent to the wall –the so-called viscous sublayer. A large body of experiments shows that in this region the diagram of the average streamwise component of the velocity of the fluid particle is linear. The same region is also known to be a site of intense turbulence. The paper shows that these two facts, which are contradictory at first sight, indicate that a sequence of rolling vortices is active in the viscous sublayer.

By applying the angular momentum balance law, we show that, in a wall-bounded flow, all vortex formation takes place in the viscous sublayer. We also determine the vortical structure of the viscous sublayer and calculate the diameter, angular velocity and interaxis of the vortices in steady-state conditions.

The vortices of the viscous sublayer reform continuously as soon as they are lifted from that sublayer to migrate toward the central regions of the flow. Vortex lifting occurs through the Magnus effect. The lift force is calculated in the paper. The spin of the vortices makes them prone to gyroscopic precession. Driven by small irregularities in the flow, gyroscopic precession reorients the vortex axis in different and unexpected directions, thus contributing to the seeming disorder of the turbulent flow.

From the conceptual standpoint, the connection between vortex production and maximum shear stress, considered in this paper, provides a rationale for the origin of turbulent vortices in viscous fluids. From a more practical standpoint, the same result should help to foster better ways –both theoretical and numerical– to model and control the turbulent flow.

REFERENCES

1.	Batchelor G.K.: *An Introduction to Fluid Dynamics*, CUP, Cambridge, UK, 2000.

2.	Panton R. L.: *Incompressible Flow*, 4$^{nd}$ ed., John Wiley & Sons, Hoboken, N.J., 2013

3.	Kundu P.K., Cohen I.M., Dowling D.R.: *Fluid Mechanics*, 5$^{th}$ ed., Academic Press, Waltham, MA, 2012.

4.	Robinson S. K.: Coherent motions in the turbulent boundary layer. *Annu. Rev. Fluid Mech.* **23**, 601-639 (1991).

5.	Guermond J.-L., Quartapelle L.: Equivalence of *u-p* and *ζ-ψ* formulations of the time-dependent Navier-

Stokes equations. *Int. J. Numer. Methods Fluids* **18**, 471-487 (1994).

6. Ern A., Guermond J.-L., Quartapelle L.: Vorticity–Velocity Formulations of the Stokes Problem in 3D. *Math. Meth. Appl. Sci.* **22**, 531-546 (1999).

7. Wu X.: Inflow turbulence generation methods. *Annu. Rev. Fluid Mech.* **49**, 23-49 (2017).

8. Kline S.J., Robinson S.K.: Turbulent Boundary Layer Structure: Progress, Status, and Challenges. In: *Structure of Turbulence and Drag Reduction* (A. Gyr ed.), *IUTAM Symp. July 25-28, 1989, Zurich, Switzerland*, pp. 3-22, Spinger-Verlag, Berlin, 1990.

9. Chow T.L.: *Classical Mechanics*, 2$^{nd}$ ed. CRC Press, Boca Raton, FL, 2013.

10. Goldstein H., Poole C.H., Safko J.: *Classical Mechanics*, 3$^{nd}$ ed. Addison-Wesley, Boston, MA, 2002

11. Kleppner D., Kolenow R. J.: *An introduction to Mechanics*. CUP, Cambridge, UK, 2010.

12. Truesdell C., Toupin R.: Classical field theories. In: *Principles of Classical Mechanics and Field Theory. Encyclopedia of Physics*, vol. **3**/1 (S. Flügge ed.), Spinger-Verlag, Berlin, 1960.

13. Aris R.: *Vectors, Tensors and The Basic Equations of Fluid Mechanics*. Dover Publications, New York, 1962.

14. Malvern L. E.: *Introduction to the Mechanics of a Continuous Medium*. Prentice-Hall, Englewood Cliffs, N.J., 1969.

15. Condiff D. W., Dahler J. S.: Fluid mechanical aspects of antisymmetric stress. *Phys. Fluids* 7, 842-854 (1954).

16. Leigh D. C.: *Nonlinear Continuum Mechanics*. McGraw-Hill, New York, 1968.

17. Paglietti A.: The role of angular momentum in the laminar motion of viscous fluids. *Continuum Mech. Thermodyn.* **29**, 611-623 (2017).

18. Paglietti A.: *Thermodynamic Limit to the Existence of Inanimate and Living Systems*, Sepco-Acerten, Milano, 2014.

19. Bakewell H.P., Lumley J.L.: Viscous Sublayer and Adjacent Wall Region in Turbulent Pipe Flow. *Phys. Fluids* **10**, 1880-1889 (1967).

20. Eckelmann H.: The structure of the viscous sublayer and the adjacent wall region in a turbulent channel flow. *J. Fluid Mech.* **65**, 439-459(1974).

21. Alfredson P.H., Johansson A.V.: The fluctuating wall-shear stress and the velocity field in the viscous sublayer. *Phys. Fluids* **31**, 1026-1033 (1988)

22. Popovich A.T., Hummel R. L.: Experimental study of the viscous sublayer in turbulent pipe flow. *AIChE J.* **13**, 854-860 (1967).

23. Escudier M.P., Poole R.J., Presti F., Dales C., Nouar C., Desaubry C., Graham L., Pullum L.: Observations of asymmetrical flow behaviour in transitional pipe flow of yield-stress and other shear-thinning liquids. *J. Non-Newtonian Fluid Mech.* **127**, 143–155 (2005).

24. Wang·L., Fu S.: Development of an intermittency equation for the modeling of the supersonic/hypersonic boundary layer flow transition. *Flow Turbulence Combust.* **87**,165–187(2011)

25. Kreplin H. P., Eckelmann H.: Behavior of the three fluctuating velocity components in the wall region of a turbulent channel flow. *Phys. Fluids* **22**, 1233-1239 (1979).

26. Meek R.L., Baer A.D.: The periodic viscous sublayer in turbulent flow. *AIChE J.* **16**, 841-848 (1970).

27. Tokumaru P. T., Dimotakis P. E.: The lift of a cylinder executing rotary motions in a uniform flow. *J. Fluid Mech.* **255**, 1-10 (1993).

28. Kang S., Choi H.: Laminar flow past a rotating circular cylinder. *Phys. Fluids* **11**, 3312-3321 (1999).

29. Stojković D., Breuer M., Durst F.: Effect of high rotation rates on the laminar flow around a circular cylinder. *Phys. Fluids* **14**, 3160-3178 (2002).

30. Çengel Y.A., Cimbala J.M.: *Fluid Mechanics: Fundamentals and Applications*, 4th ed., McGraw-Hill, Boston, Mass., 2017.

31. Cheng M., Luo L.S.: Characteristics of two-dimensional flow around a rotating circular cylinder near a plane wall. *Phys. Fluids* **19**, 063601 1-17 (2007).

32. Rao A., Thompson M.C., Leweke T., Hourigan K.: Flow past a rotating cylinder translating at different gap heights along a wall. *J. Fluids Struct.* **57**, 314–330 (2015).

33. Praturi A.K., Brodkey R. S.: A stereoscopic visual study of coherent structures in turbulent shear flow. *J. Fluid Mech.* **89**, 251-272 (1978).

34. Wu Y., Christensen K. T.: Population trends of spanwise vortices in wall turbulence. *J. Fluid Mech.* **568**, 55-76 (2006).